\documentclass[fleqn,usenatbib,useAMS]{mnras}

\usepackage{graphicx}	% Including figure files
\usepackage{amsmath}	% Advanced maths commands
\usepackage{amssymb}	% Extra maths symbols
\usepackage{multicol}        % Multi-column entries in tables
\usepackage{multirow}
\usepackage{bm}		% Bold maths symbols, including upright Greek
\usepackage{pdflscape}	% Landscape pages

\usepackage{xspace}
\usepackage[capitalize]{cleveref}
\usepackage[dvipsnames]{xcolor}
\usepackage{siunitx}

\usepackage[T1]{fontenc}
\usepackage{ae,aecompl}
\usepackage{newtxtext,newtxmath}

\pubyear{2021}

\newcommand{\prx}{Phys. Rev. X}

\newcommand{\Mchirp}{\mathcal{M}}
\newcommand{\phiJL}{\phi_{\rm JL}}
\newcommand{\dL}{d_{\rm L}}
\newcommand{\tgeo}{t_{\rm geo}}
\newcommand{\thetaJN}{\theta_{\rm JN}}
\newcommand{\order}[1]{\ensuremath{\mathcal{O}}(#1)\xspace}

\newcommand{\IMRPhenomPvTwoNRTidal}{{\texttt{IMRPhenomPv2\_NRTidal}}\xspace}
\newcommand{\IMRPhenomPvTwo}{{\texttt{IMRPhenomPv2}}\xspace}
\newcommand{\TaylorF}{{\texttt{TaylorF2}}\xspace}

\newcommand{\lowspin}{\ensuremath{a_i{<}0.05}\xspace}
\newcommand{\zerospin}{\ensuremath{a_i{=}0}\xspace}
\newcommand{\zerospinequalmass}{\ensuremath{a_i{=}0,q{=}1}\xspace}

\newcommand{\bilby}{\texttt{bilby}\xspace}
\newcommand{\dynesty}{\texttt{dynesty}\xspace}
\newcommand{\bayestar}{\texttt{BAYESTAR}\xspace}
\newcommand{\LALInference}{\texttt{LALInference}\xspace}
\newcommand{\RIFT}{\texttt{RIFT}\xspace}
\newcommand{\scipy}{\texttt{scipy}\xspace}

\begin{document}
\label{firstpage}
\pagerange{\pageref{firstpage}--\pageref{lastpage}}

%EHT:
%\title[Optimized source localisation]{Optimized source localisation for gravitational-wave binary coalescence events}
\title[Optimized source localisation]{Optimized localisation for gravitational-waves from merging binaries}
\author[You et al.]{Zhi-Qiang You$^{1,2,3,6}$, Gregory Ashton$^{2,3,4}$ \thanks{E-mail: gregory.ashton@ligo.org}, Xing-Jiang Zhu$^{2,3,6}$, Eric Thrane$^{2,3}$,
\newauthor{and Zong-Hong Zhu$^{1,5,6}$}
\\
$^{1}$School of Physics and Technology, Wuhan University, Wuhan 430072, China\\
$^{2}$School of Physics and Astronomy, Monash University, Clayton, Vic 3800, Australia\\
$^{3}$OzGrav: The ARC Centre of Excellence for Gravitational Wave Discovery, Clayton, Vic 3800, Australia\\
$^{4}$Department of Physics, Royal Holloway, University of London, TW20 0EX, United Kingdom\\
$^{5}$Department of Astronomy, Beijing Normal University, Beijing 100875, China\\
$^{6}$Advanced Institute of Natural Sciences, Beijing Normal University at Zhuhai 519087, China
}
\maketitle
\begin{abstract}
The Advanced LIGO and Virgo gravitational wave observatories have opened a new window with which to study the inspiral and mergers of binary compact objects. These observations are most powerful when coordinated with multi-messenger observations. This was underlined by the first observation of a binary neutron star merger GW170817, coincident with a short Gamma-ray burst, GRB170817A, and the identification of the host galaxy NGC~4993 from the optical counterpart AT~2017gfo. Finding the fast-fading optical counterpart critically depends on the rapid production of a sky-map based on LIGO/Virgo data. Currently, a rapid initial sky map is produced followed by a more accurate, high-latency, $\gtrsim\SI{12}{hr}$ sky map.
We study optimization choices of the Bayesian prior and signal model which can be used alongside other approaches such as reduced order quadrature. We find these yield up to a $60\%$ reduction in the time required to produce the high-latency localisation for binary neutron star mergers.
\end{abstract}


\section{Introduction}
\label{sec:introduction}
Transient multi-messenger astronomy is delivering new insights into the nature of compact objects, the cosmological properties of our Universe, and general relativity. During their second observing run (O2), the the Advanced-LIGO \citep{2015CQGra..32g4001L} and Virgo \citep{2015CQGra..32b4001A} detectors observed gravitational-wave emission from a binary neutron star merger, GW170817 \citep{abbott17_gw170817_detection}. This observation coincided with a short gamma-ray burst, GRB170817A \citep{goldstein2017ordinary} and follow-up campaigns across the electromagnetic spectrum \citep{abbott17_gw170817_multimessenger} led to the observation of an optical counterpart AT2017gfo \citep{valenti2017discovery,yang2017empirical,perego20172017gfo} and identification of the host galaxy NGC 4993. This multi-messenger view of the event provided critical insights into multi-messenger astronomy, opening a new path by which to study and understand the mergers of neutron star binaries, short gamma-ray bursts, and their optical counterparts.

The LIGO and Virgo observatories, joined by KAGRA \citep{kagra}, have now completed their third observing run \citep{gwtc2_review}. During this observing run, open public alerts were issued (see \href{https://emfollow.docs.ligo.org/userguide/}{emfollow.docs.ligo.org}) enabling numerous follow-up campaigns (see, e.g., \citet{2020MNRAS.497.5518A, 2020A&A...643A.113A, 2019ApJ...885L..19C, 2020MNRAS.497..726G}.
So far, there has yet to be a bona-fide multi-messenger observation from the O3 observing run [however, see \citet{graham2020, Ashton2020} for discussion of a speculative connection between the binary black hole merger GW190521~\citep{GW190521} and an active galactic nucleus flare].
In preparation for the fourth observing run (estimated to be June 2022), the detectors are currently being upgraded with a projected binary neutron star inspiral range of \SI{190}{Mpc} \citep{whitepaper}.
During this observing run, the Advanced-LIGO, Virgo, and KAGRA (HLVK) network is likely to observe tens of transient systems containing a neutron star and hundreds of binary black hole systems.
Binary neutron star (BNS) mergers are known to produce a multi-messenger counterparts~\citep{abbott17_gw170817_multimessenger} known as a ``kilonova'' \citep{li1998}.
As yet, no unambiguous detection of a neutron-star black-hole (NSBH) binary has been made; but such systems may also produce an electromagnetic counterpart \citep{lattimer1974, li1998} (see also \citet{fernandez2016} for a review).
As such, mergers containing a neutron star tend to be prioritised by follow-up campaigns,
however, as demonstrated by \citet{graham2020}, the next big surprise may instead come from a binary black hole (BBH) merger.

Online gravitational-wave searches are able to identify events in the data and analyse their significance in a timescale of \order{s}. Following the identification, the event is automatically vetted and published in a GCN notification in a timescale of \order{\SI{10}{s}}. Optical telescopes then perform (often automated) searches for the rapidly-fading, timescales of \order{hrs}, electromagnetic transient. 
The ability to identify the transient is highly dependent on the three-dimensional (3D) source-localisation [here, we parameterise in terms of the right-ascension (RA), declination (DEC), and luminosity distance $d_{L}$]. 
In the first three observing runs, results from two methods for producing localisations were published.
First, the \emph{low-latency} \bayestar \citep{singer16a,singer16b} algorithm produces a 3D localisation in \order{minutes} utilizing the maximum-likelihood template from the search pipelines.
Second, the \emph{high-latency} {\tt{LALInference}} \citep{veitch15} algorithm, which uses stochastic sampling to construct the full posterior probability distribution and produces results in a timescale of \order{>12 hrs}.
While for the majority of systems, differences between the low-latency and high-latency localisation are anticipated to be small (\cite{LVK_localization}), the high-latency localisation is preferable since it can include an improved physical description of the signal and noise. Moreover, a systematic study (\cite{morisaki2020}) comparing the \bayestar algorithm with a stochastic-sampling based approach (discussed below), demonstrates that \bayestar can over-estimate the localisation uncertainty when the best-fit parameters of the signal are outside the online-detection-pipeline template bank. This can happen, for example, if the online pipeline does not include the spin-precession of the source. 

Constructing the full posterior probability distribution is a computationally challenging task. Stochastic sampling methods such as Markov-Chain Monte-Carlo (MCMC) \citep{metropolis1953equation, hastings1970monte} and Nested Sampling \citep{Skilling06}, require $\sim 10^{6}-10^{8}$ evaluations of the likelihood to analyse gravitational-wave signals \citep{christensen1998, veitch08}.
Typically, each evaluation of the likelihood is dominated by the time required to model the source.
This time varies between a few milliseconds to several tens of seconds depending on the signal duration and sophistication of the waveform model. 
As such, the \emph{wall time} required to draw a sufficient number of samples to approximate the posterior can be between several hours, to many tens of days.
A number of advances have been made in reducing the wall time:
\begin{enumerate}
\item
\textit{Reduced-order-quadrature (ROQ) methods} \citep{2012arXiv1210.0577A, 2013PhRvD..87l4005C, 2015PhRvL.114g1104C, smith16, 2021PhRvD.104f3031Q} interpolate the likelihood to high accuracy and can speed up evaluation times by factors of several hundred.
While ROQ-based method have enjoyed considerable success \citep[see, e.g.,][]{abbott20_GW190425}, they do require that the ROQ basis be pre-constructed, often at significant computational cost.
As such, their utility can be limited for online production of the 3D localisation if the pre-computed basis set does not cover the required parameter space.
\cite{morisaki2020} recently demonstrated that so-called focused-ROQ (FROQ), in which many bases covering narrow ranges of the parameter space offer greater speed-ups still: with gains of up to $10^{4}$ seen for low-mass systems.

\item
\emph{Hetrodyned likelihoods} \citep{2010arXiv1007.4820C, 2021PhRvD.103j4057C}, also known as the \emph{relative-binning} method \citep{zackay18, daniel_RB_fast},
exploit the computation for likelihoods of similar waveforms, whose phases and amplitudes differ smoothly with frequency, by pre-computing frequency–binned overlaps of the best–fit waveform with the data.
These methods do not require a pre-computation step and offer speed-ups of up to $\sim$ $10^4$ in likelihood evaluation times.
The accuracy of these approaches depends on the expansion order: just a few terms are required to sufficiently approximate the likelihood.
Demonstrations of this method are very promising, but work is needed to verify the accuracy and limitations of the method against the full likelihood.

\item Then, there is \textit{brute-force parallelisation}. Parallelisation can be done at the level of the likelihood itself \citep{talbot19}, the stochastic sampler, or using multiple independent stochastic samplers. The later two aspects have been generously employed in standard inference packages \citep{veitch15, ashton19, biwer18} using the few tens of cores available on typical central processing units, while \cite{Smith:2019ucc} demonstrated the capacity to scale to the many hundreds of cores available in high-performance computers using the \dynesty \citep{dynesty} nested sampling algorithm.
Such approaches are useful as they do not require pre-computation and make no requirements about the waveform itself.
Unlike the other techniques to reduce wall time, this technique is not ``free'' (achieved through clever design)---it requires additional computing resources.

\item The \RIFT family of stochastic samplers \citep{2015PhRvD..92b3002P, 2018arXiv180510457L} employs aspects of brute-force parallelisation (with extensions to graphical processing units \citep{2019PhRvD..99h4026W}) alongside pre-computing aspects of the waveform in order to carry out inference with iterative fitting.

\item Significant speed-ups may be realised by the use of machine-learning based approaches. Such approaches do not apply the standard principles of stochastic sampling; instead, the algorithm is pre-trained on example of signals in noise and can then produce posterior samples within a few seconds \citep{gabbard19, green2020, green2021complete}.
Such algorithms present a significant opportunity as they could handle non-Gaussian noise and arbitrarily complex waveforms by developing realistic training sets.
For these approaches, the optimizations discussed herein are not directly applicable, in the sense that they do not consider a specific prior or waveform model during the analysis.
However, the lessons learned can be applied in selecting the complexity of training data.

\item Finally, there are many other approaches to speeding up stochastic sampling such as the use of machine-learning coupled nested sampling \citep{2021PhRvD.103j3006W}, adaptive frequency-banding \citep{2021PhRvD.104d4062M}, and representing the signals in the time-frequency domain \citep{2020PhRvD.102l4038C}.

\end{enumerate}

In the fourth observing run, one (or many) of these approaches may be used to reduce the latency of full parameter-estimation results.
However, the wall time they require (or the training time in the case of machine-learning based approaches) still depends on the choice of waveform model and the astrophysical prior.
In this work, we study how to optimize the choice of model and prior to reduce the wall time.
We will develop these ideas in the context of ROQ-methods and parallelisation, but the ideas apply equally to many of the methods listed above.

This paper is organized as follows. In \cref{sec:optimized}, we describe the optimization of the prior then in \cref{sec:result}, we validate which of these optimizations produces acceptably small bias and calculate the improvement in wall time. In \cref{sec:waveform}, we discuss optimization of the waveform model.
In \cref{sec:summary}, we conclude with a discussion on how these choices can be used during the next observing run to minimize the time to produce localisation.

\section{Optimized choices for priors} \label{sec:optimized}

A compact binary coalescence (CBC) signal is described by up to 10 intrinsic parameters and 7 extrinsic parameters. The intrinsic parameters are: chirp mass $\Mchirp$, mass ratio $q$, component spin magnitude ($a_1$ for the primary heavier-mass object and $a_2$ for the secondary lighter-mass object), and four angles describing the spin orientation ($\theta_1, \theta_2, \phi_{12}$, and $\phiJL$). For systems containing one or more neutron stars, the intrinsic parameters also include the neutron star tidal deformability $\Lambda_1$ and $\Lambda_2$.
The extrinsic parameters are: the 3D localisation coordinates (RA, DEC, $\dL$), the polarization angle of the source $\psi$, the GPS reference time $\tgeo$ of the merger, the angle between the total angular momentum and the line of sight $\thetaJN$, and the binary phase $\phi$ (defined at a fixed reference frequency). (See Table~E1 of \citet{Romero_2020_bilby_gwtc} for a detailed description of these parameter and alternative parameterisations).\footnote{
In addition, there may by up to $20$ parameters per-detector to describe the detector calibration (see \citet{cahillane17} for a description, \citet{Romero_2020_bilby_gwtc} for a discussion of how these are marginalized in the \bilby software used herein, and \citet{2020PhRvD.102l2004P, Vitale2021} for new more physical approaches to marginalizing calibration uncertainty).
However, \citet{2020PhRvD.102l2004P} suggest that the effect of calibration uncertainty is likely to be negligible during the advanced-detector era, and that calibration uncertainty can be added to inference calculations in post-processing.
}
For stochastic samplers, the number of likelihood evaluations (and hence the wall time) depends on the complexity of the posterior.
As a general rule of thumb, the number of parameters is a good leading-order description of the complexity: more parameters require more likelihood evaluations and hence a longer wall time.
Optimizing the choice of prior can help minimize the wall time while ensuring the result remains unbiased.

First, we discuss different options for the spin prior. In an \emph{aligned-spin} prior, the source model is restricted to solutions excluding precession \citep{2015PhRvD..91b4043S}. In a \emph{low-spin} model, the magnitudes of the spin are restricted (typically to a dimensionless spin of $0.05$ \citep{abbott18_GW170817_NS_parameters}). 
The upper end of the low-spin prior corresponds to a conservative limit on the effective spins of pulsars in known Galactic double neutron stars that are capable of merging within a Hubble time \citep{Zhu-Ashton20}.
Further, one can neglect the effects of spin entirely. It was shown by \citet{farr2016parameter} that the 3D localisation for low-spin binary neutron star (BNS) sources is unbiased given a zero-spin prior.
However, black-hole binaries can sustain significant dimensionless spin (see, e.g., \cite{gwtc2_population}).

Second, we examine the priors for tidal parameters.
For BNS and NSBH binaries, the tidal parameter are relatively poorly measured \citep{abbott18_GW170817_NS_parameters, abbott20_GW190425}. A straightforward prior choice is to neglect tidal parameters (in effect, we set $\Lambda_1=0$ and $\Lambda_2=0$, i.e. that both components are black holes).
The logic here is that, if we cannot accurately measure the tides, the tidal parameters probably do not have a strong affect on the sky localisation.

Third, we consider the prior for mass ratio.
For the two confident BNS events in Gravitational-wave Transient Catalog 2 (GWTC-2; \cite{gwtc2_review}), the masses of the two components are nearly equal.
Under the astrophysical low-spin prior, the mass ratio is constrained to be between $0.7-1$ and $0.8-1$ for GW170817 and GW190425 respectively \citep{abbott18_GW170817_NS_parameters, abbott20_GW190425}. This suggests another potential prior optimization: restricting to equal-mass systems $q=1$.

We investigate these optimization strategies. In \cref{tab:priors}, we list four priors distributions.
A precessing, but low-spin \emph{astrophysical prior}, captures our broad expectation for the typical population parameters of neutron star binaries; it is from this prior that we draw simulation parameters.
The remaining three priors combining increasing restrictions on the spin and mass-ratio.
For all three choices, we assume zero tidal deformability: this prior optimization was already used during the third observing run for high-latency 3D localisation.

\begin{table}\label{tab:prior_model}
    \centering
    \caption{
    Definition of the astrophysical, low-spin (\lowspin), zero-spin (\zerospin), and zero-spin and equal-mass (\zerospinequalmass) priors investigated in this work.
    We define $a_i$ and $\Lambda_i$ to be the absolute spin magnitude and tidal deformability of the $i^{\rm th}$ component in the binary and $q\le 1$ to be the mass ratio.
    For the astrophysical prior and low-spin prior, the spins are fully-precessing but restricted in magnitude. For all other parameters, the priors follow the standard distributions described in Table E1 of \citep{Romero_2020_bilby_gwtc}.
    \label{tab:priors}}
    \begin{tabular}{l c c c c}
    \hline
    \hline
    \  & astrophysical & \lowspin & \zerospin & \zerospinequalmass \\
    \hline
          $a_1$ & $[0, 0.05]$ & $[0, 0.05]$ & $0$ &  $0$\\
          $a_2$ & $[0, 0.05]$ & $[0, 0.05]$ & $0$ &  $0$\\
          $q$  &  $[0.125, 1.0]$ &  $[0.125, 1]$ & $[0.125, 1]$ & $1$ \\
          $\Lambda_1$ & $[0, 5000]$ & $0$ & $0$ &  $0$\\
          $\Lambda_2$ & $[0, 5000]$ & $0$ & $0$ &  $0$\\
    \hline
    \hline
    \end{tabular}
\end{table}

\section{Validation of prior optimization} \label{sec:result}

The \lowspin, \zerospin, and \zerospinequalmass priors defined in \cref{tab:priors} increasingly constrain the astrophysical properties of the source.
Almost certainly, these over-constrain the source properties. For example, GW190425 shows some support for spin effects and is unlikely to be equal mass while the data from GW170817 does constrain the tidal deformability.
The question is, do these unphysical prior constraints bias the 3D sky localisation? To answer this question, we perform tests on simulated data and look at the localisation of detections during previous observing runs that have the potential for electromagnetic counterparts.

\subsection{Parameter-parameter test}
\label{sec:simulated}

For our first test, we simulate 100 binary neutron stars using the \IMRPhenomPvTwoNRTidal waveform \citep{Dietrich19}. The simulated signals have spin magnitudes, mass ratios, and tidal deformability parameters drawn from the astrophysical prior (see \cref{tab:priors}).
For the remaining parameters we draw them from the standard astrophysical distributions.

The signals are simulated in \SI{128}{s} of data using the projected O4 sensitivity colored Gaussian noise \citep{whitepaper} for the HLV detector network.
Signals with a network signal to noise ratio (SNR) less than 12 are discarded, with replacement; this ensures the injection set reflects the population of events from which we are likely to observe electromagnetic counterparts: high-SNR systems with good sky localisation.
We analyse the simulated data sets using the \IMRPhenomPvTwo \citep{IMRPhenomP,hannam2014,khan16} waveform model (which excludes tidal deformability), an ROQ-basis \citep{smith16}, and the \dynesty nested sampling algorithm as implemented in \bilby \citep{Ashton2020}.
The ROQ basis is limited to chirp mass values of $1.42$ to $2.60$ solar masses; as such, we limit the prior distribution on chirp mass (and similarly the distribution from which we draw simulation parameters) to this range.

\begin{figure}
\centering
\includegraphics[width=.48\textwidth]{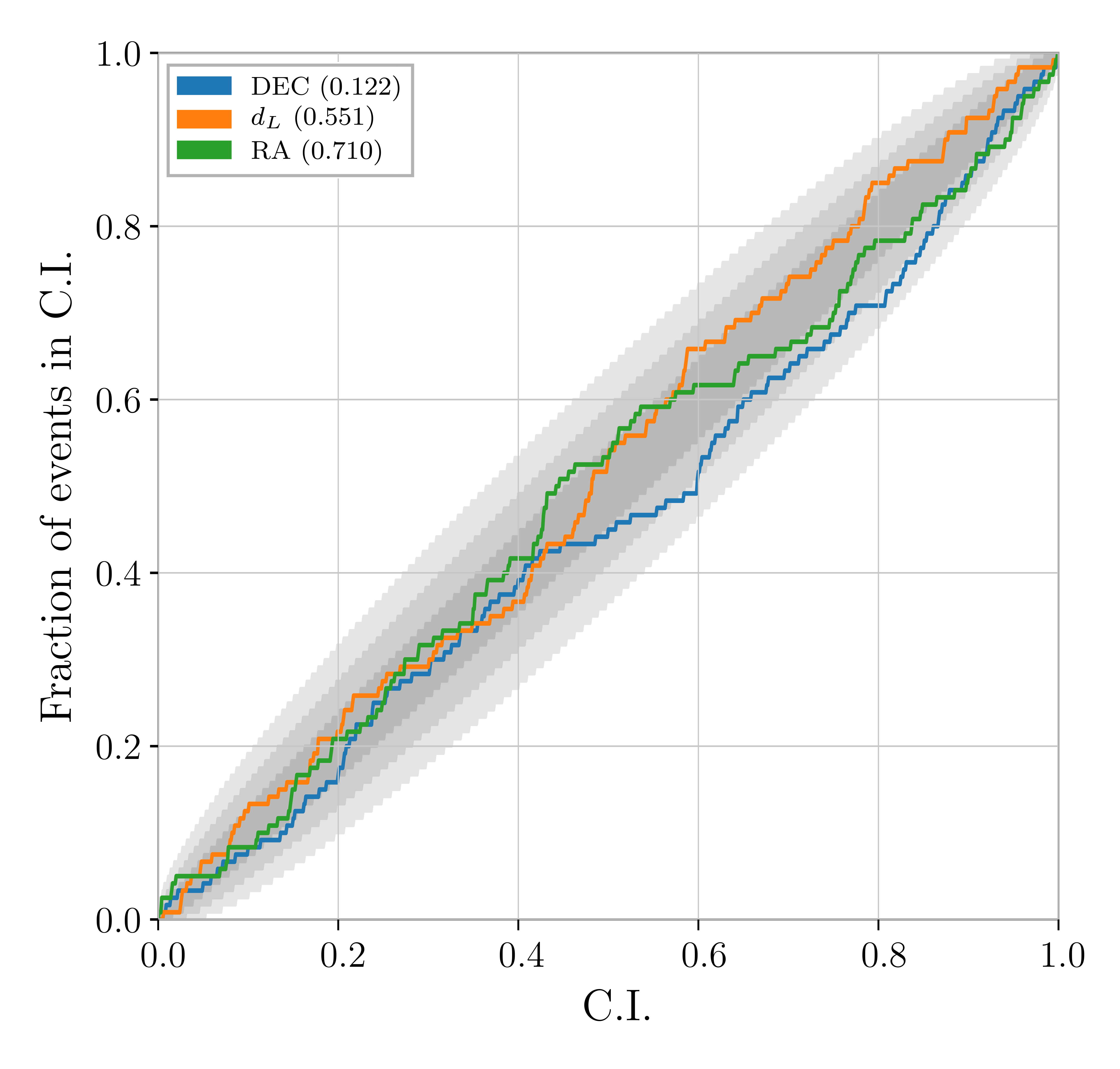}
\caption{A parameter-parameter (PP) test of the 3D sky localisation parameters of 100 simulated BNS signal using the \zerospinequalmass prior. We simulate signals drawn from the astrophysical prior \cref{tab:priors} using the \IMRPhenomPvTwoNRTidal waveform model and analyse the data using the \IMRPhenomPvTwo waveform model. The grey region indicates the 1, 2, and 3$\sigma$ confidence intervals. Individual p-values for each of the 3D localisation parameters are shown in legend.}
\label{fig:pp_0S1Q}
\end{figure}

We find that all choices of prior specified in \cref{tab:prior_model}, the 3D localisation is unbiased. We test this using a PP, or parameter-parameter, test \citep{cook06, talts18} over the 100 simulated signals.
In \cref{fig:pp_0S1Q}, we show the results of the PP test for the most restrictive zero-spin equal-mass (\zerospinequalmass) prior.
Qualitatively, bias manifests in a PP plot as a deviation in the parameter curve from the diagonal: for all parameters in \cref{fig:pp_0S1Q}, the curves remain inside the 3$\sigma$ uncertainty region.
One way to quantify the bias from a PP plot is to calculate the p-value expressing the probability that the fraction of events in a particular confidence interval is drawn from a uniform distribution. 
We calculate the p-value using the Kolmogorov-Smirnov test as implemented in \scipy \citep{scipy}.
For the most restrictive \zerospinequalmass prior the p-values are 0.122, 0.551, and 0.710 for DEC, $d_L$, and RA, respectively.
That none of these $p$-values are small $\ll 0.05$ indicates that there is no measurable sign of bias.

As a further way to understand the bias, in \cref{tab:out_fraction}, we calculate the fraction of events outside of the 90\% and 68\% credible intervals for each prior.
For an unbiased inference, we would expect the fraction out of the 90\% and 68\% credible intervals to be 10\% and 32\% respectively.
The difference then provides a quantitative estimate of the bias. For the \lowspin prior, this is up to 2\% while for the \zerospinequalmass prior, it is up to 11\%.

In Appendix \ref{sec:append}, we show additional two PP plots for the \zerospinequalmass prior for a two-detector network (Fig. \ref{fig:pp_2det}) and for 500 injections (Fig. \ref{fig:pp_500}). Furthermore, in Table \ref{tab:comp_n}, we list the p-values and fractions of truth parameters that are out of the 90\% credible intervals, for 50, 100, 300 and 500 injections.
These illustrate that our results are robust against the choice of detector network configuration and the number of injections.

\begin{table}
    \centering
    \caption{The fraction of the truth value out of 90\%, 68\% credible interval for \lowspin and \zerospinequalmass priors for the 100 simulated BNS discussed in \cref{sec:simulated}.
    \label{tab:out_fraction}}
    \begin{tabular}{l|cccc}
    \hline
     Model & Fraction & $\rm{DEC}$ & $\rm{RA}$ & $d_L$ (Mpc) \\
    \hline\hline
       \multirow{2}{*}{\lowspin} & out-of-90\%  & 9\% & 12\%  & 10\% \\
        & out-of-68\% & 32\% & 33\% & 31\% \\
    \hline
       \multirow{2}{*}{\zerospinequalmass} & out-of-90\%  & 13\% & 12\%  & 9\% \\
        & out-of-68\% & 37\% & 35\% & 21\% \\
    \hline
    \end{tabular} 
\end{table}

\begin{figure}
\centering
\includegraphics[width=.5\textwidth]{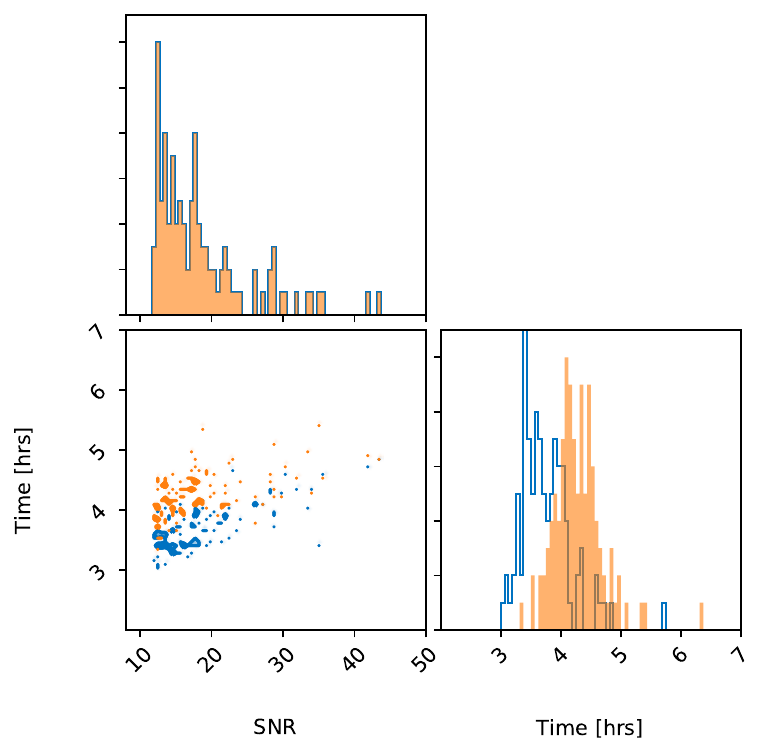}
\caption{The SNR of simulated signals and the measured wall time in hours for the 100 simulated BNS events obtained using the \zerospinequalmass (blue) and \lowspin (orange) priors studied in \cref{sec:simulated}.
All wall-times are evaluated on an Intel Gold 6140 CPU processors using 32 threads.}
\label{fig:snrtime}
\end{figure}

To compare the wall time between the \lowspin and \zerospinequalmass model, in \cref{fig:snrtime}, we plot the optimal SNR of the simulated signals and the sampling time.
As expected for a nested sampling algorithm, the sampling time is correlated with the signal SNR. (The sampling time is generally proportional to the ratio of prior to posterior volume \citep{dynesty}; at a fixed prior volume, the posterior volume decreases as the SNR increases leading to longer wall times).
The mean sampling time for the \lowspin and \zerospinequalmass priors are $\sim$ 4.5 hours and 3.5 respectively. This demonstrates that, in addition to the low-spin savings already identified by \citet{farr2016parameter}, utilising an equal-mass prior can provide a further $\sim20\%$ performance improvement. 

\subsection{Implications for BNS and NSBH localisation}
\label{sec:bns-follow-up}

To date, the LIGO and Virgo detectors have observed two BNS events, GW170817 \citep{abbott17_gw170817_detection} and GW190425 \citep{abbott20_GW190425},
and two NSBH events, GW200105 and GW200115 \citep{LVKnsbh21}.
These systems are likely to be accompanied by a rapidly fading electromagnetic counterpart and are hence prioritised for follow-up by optical telescopes \citep{fernandez2016}.
As such, we have the most to gain in optimizing the production of the high-latency 3D localisation.

\begin{figure}
\centering
\includegraphics[width=.5\textwidth]{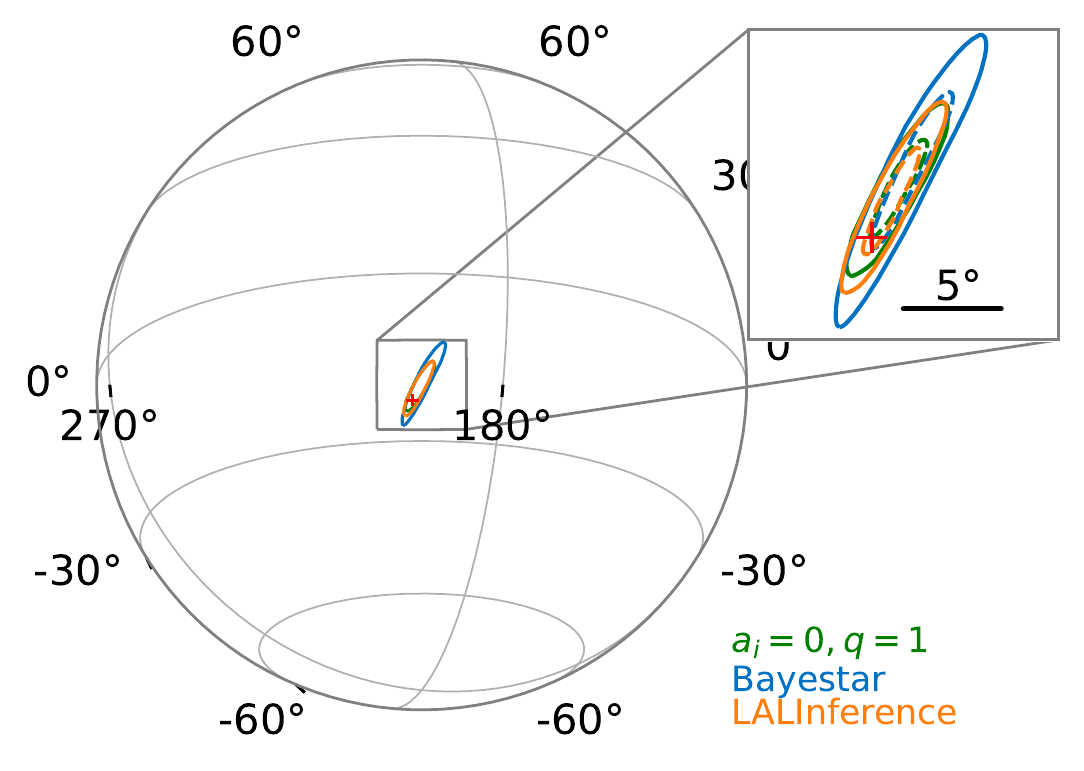}
\caption{Sky location uncertainty for GW170817. We show posteriors from the \zerospinequalmass prior in green, the O3 low-latency \bayestar analysis in blue, and the O3 high-latency \LALInference analysis in orange. The red ``plus'' is the location of the identified host galaxy \citep{abbott17_gw170817_multimessenger}.
The solid (dashed) line represents the 90\% (50\%) credible region.
}
\label{fig:GW170817}
\end{figure}

\begin{figure}
\centering
\includegraphics[width=.5\textwidth]{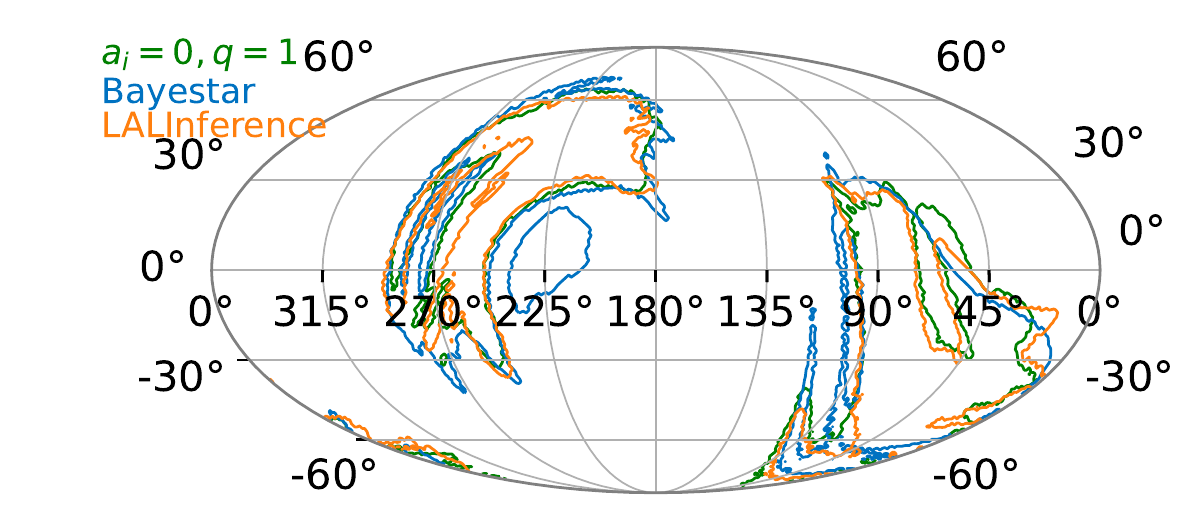}
\caption{Sky location uncertainty for GW190425. We show posteriors from the \zerospinequalmass prior in green, the O3 low-latency \bayestar analysis in blue, and the O3 high-latency \LALInference analysis in orange. 
The solid line represents the 90\% credible region.
}
\label{fig:GW190425}
\end{figure}

To study our optimized priors for BNS signals, we re-analyse the public data \citep{gwosc} for the BNS events GW170817 and GW190425 using the \zerospinequalmass prior and the \IMRPhenomPvTwo waveform model.
In \cref{fig:GW170817} and \cref{fig:GW190425}, we plot the sky localisation uncertainty from our reanalysis, the O3 low-latency (\bayestar), and the O3 high-latency (\LALInference) results for GW170817 and GW190425, respectively.

High-latency results better-constrain the sky-localisation: this can be seen in \cref{tab:credible-interval} where we report the area coverage and was demonstrated systematically by \citet{morisaki2020}.
\cref{tab:credible-interval} demonstrates that the \zerospinequalmass prior provides an equivalent constraint on the sky area to the O3 high-latency results for BNS systems. We conclude then that for BNS, any method attempting to reduce the wall time of high-latency localisation results (e.g. the FROQ method \cite{morisaki2020}), could further gain from an optimized choice of prior.

The obvious concern in using an optimized prior with zero-spin and equal-mass is the induced bias if the detectors observe an event which is highly spinning or of unequal mass ratio.
We have controlled for this, to some extent, in our PP tests (\cref{sec:simulated}) by allowing our simulated signals to be drawn from the ``astrophysical prior'' (cf. \cref{tab:priors}).
In particular, we consider mass ratios down to $1/8$.
In comparison, the mass ratios for two recent NSBH events are around $1/5$ \citep{LVKnsbh21}.
The spin measurements are dominated by the contribution from the more massive BHs: whereas the dimensionless BH spin magnitude $a_1$ is constrained to less than 0.23 in GW200105 at the 90\% credible level, $a_1$ could be as high as $0.5$ (though with a large uncertainty and still being consistent with zero) for GW200115 \citep{LVKnsbh21}.

This suggests the \zerospinequalmass prior may be appropriate for NSBH systems, provided the mass ratio is not too extreme and that the black hole does not have significant spin.
Nevertheless, a real signal which falls outside of our astrophysical prior may be one of the most exciting opportunities for electromagnetic follow-up.
As such, while proposing the use of the \zerospinequalmass prior, we also recommend parallel computations be made with a broad non-informative prior.
In most instances this will lead to a small amount of wasted computation as the localisation will be essentially identical. But, this provides a safeguard against ``over-optimization.''

\begin{table*} 
    \centering
    \begin{tabular}{lccccc}
    \hline 
    \hline
     & \text{Parameters}& \zerospinequalmass & \zerospin & O3 low-latency & O3 high-latency\\ 
    \hline
    \multirow{2}{*}{\text{GW170817 (BNS)}} & $d_L\rm{(Mpc)}$ & $39^{+6}_{-9}$ & $-$ & $40^{+8}_{-8}$ & $40^{+8}_{-14}$\\
    &$\rm{Area(deg^2)}$ & $16$& $-$ & $31$  &$16$ \\
    \hline
    \multirow{2}{*}{\text{GW190425 (BNS)}}&$d_L\rm{(Mpc)}$ & $158^{+43}_{-46}$& $-$ & $155^{+45}_{-45}$ & $156^{+41}_{-41}$\\
    &$\rm{Area(deg^2)}$ &$8012$ & $-$ & $10183$ &$7461$ \\
    \hline
    \multirow{2}{*}{\text{GW190814 (BBH/NSBH)}}&$d_L\rm{(Mpc)}$& $-$& $276^{+54}_{-69}$& $236^{+53}_{-53}$&$267^{+52}_{-52}$\\
    &$\rm{Area(deg^2)}$ & $-$& $37$ &$38$  &$23$ \\
    \hline
    \multirow{2}{*}{\text{GW190412 (BBH)}}&$d_L\rm{(Mpc)}$& $-$& $738^{+237}_{-211}$& $812^{+194}_{-194}$& $734^{+138}_{-173}$\\
    &$\rm{Area(deg^2)}$ & $-$& $142$ & $156$ &$110$ \\
    \hline
    \end{tabular}

    \caption{The 68\% credible intervals for the luminosity distance ($d_{L}$) and sky position from our optimized priors, the publicly available O3 low-latency software (\bayestar), and O3 high-latency software (\LALInference). Below the event name, we give the credible source classification. For BNS events, we use the optimized \zerospinequalmass prior while we use the \zerospin for events containing a black hole. We note that some of the reduction in the localisation of between the low-latency and high-latency results for GW170817 arises from recalibration of the data from the Virgo interferometer \citep{abbott18_GW170817_properties}.}
    \label{tab:credible-interval}
\end{table*}

\subsection{Implications for BBH localisation}
Having studied prior-optimization for the localisation of BNS and NSBH events, we now turn to BBH events. BBH systems have significant spin \citep{gwtc2_population} and have been observed with highly asymmetric masses \citep{abbott2020gw190814, GW190521}.
This suggests the optimized priors may perform poorly.
To test this, we apply the \zerospin prior to the public data \citep{gwosc} for the GW190814 \citep{abbott2020gw190814} and GW190412 \citep{abbott20_GW190412} events. We report the credible intervals in \cref{tab:credible-interval}.
For both events, we find that the \zerospin prior constrains the posterior to a region almost as large as the O3 low-latency results.

\begin{figure}
\centering
\includegraphics[width=0.5\textwidth]{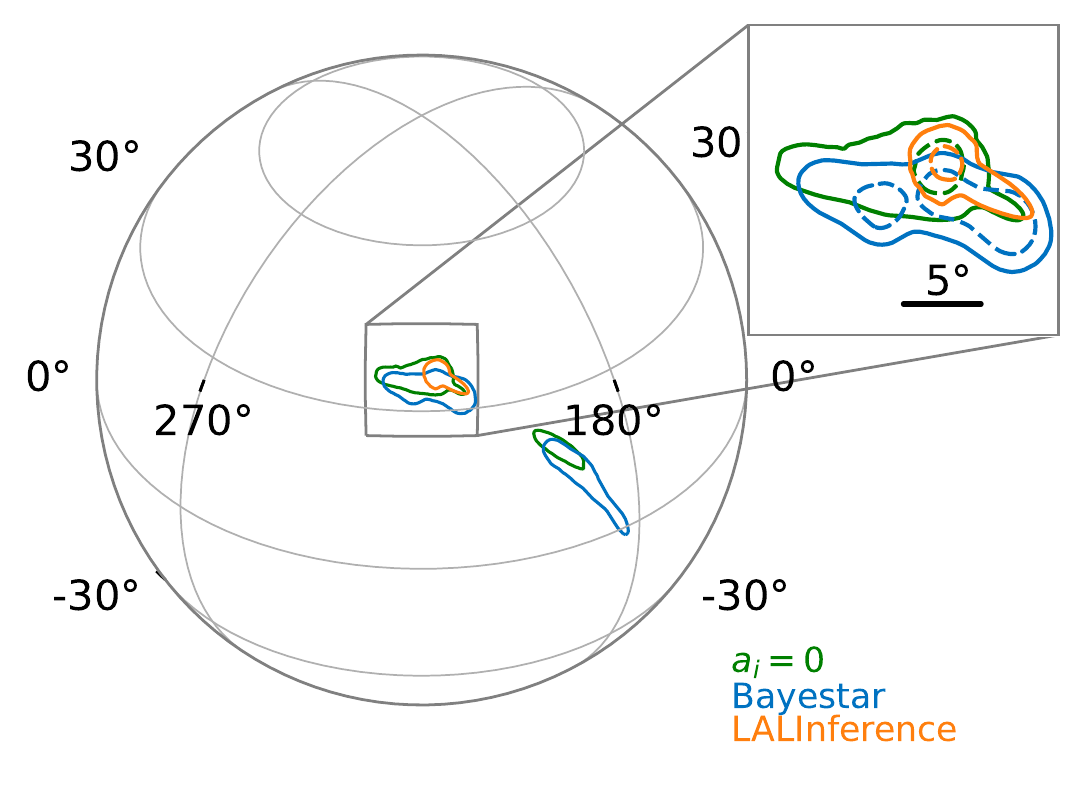}
\caption{Sky location uncertainty for GW190412 \citep{abbott20_GW190412}. We show posteriors from the \zerospin prior in green, the O3 low-latency \bayestar analysis in blue, and the O3 high-latency \LALInference analysis in orange.
The solid (dashed) line represents the 90\% (50\%) credible region.
}
\label{fig:GW190412}
\end{figure}

\begin{figure}
\centering
\includegraphics[width=0.5\textwidth]{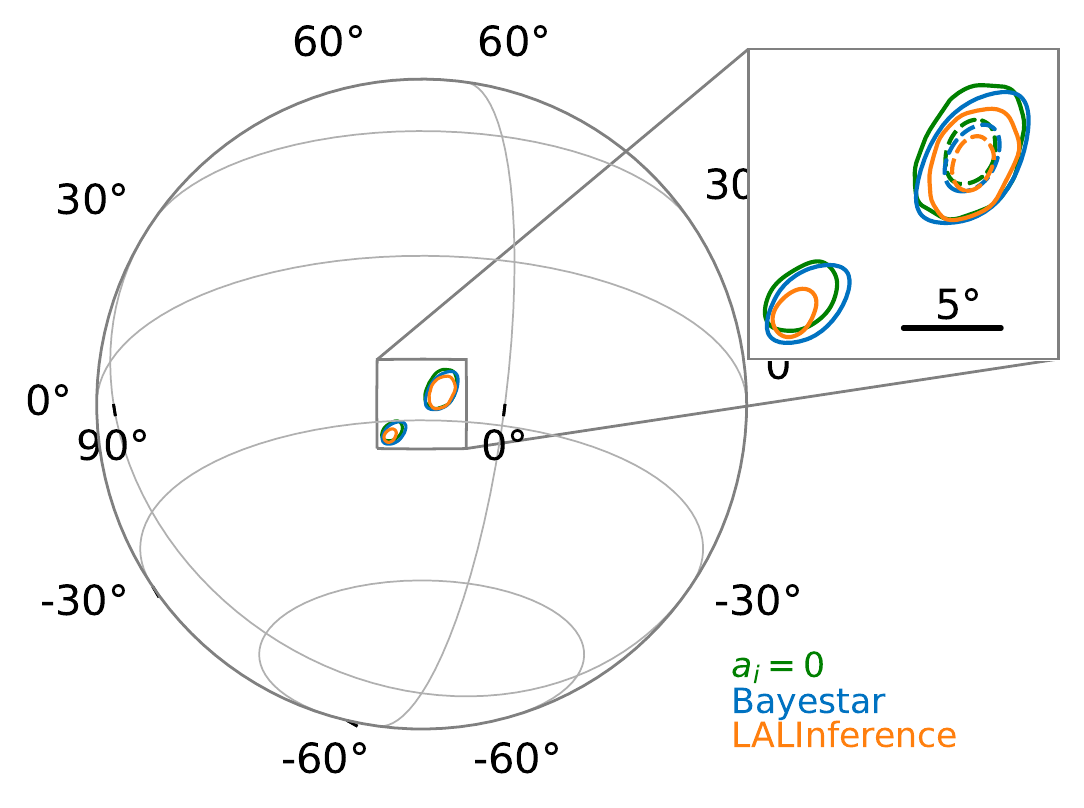}
\caption{Sky location uncertainty for GW190814 \citep{abbott2020gw190814}. We show posteriors from the \zerospin prior in green, the O3 low-latency \bayestar analysis in blue, and the O3 high-latency \LALInference analysis in orange.
The solid (dashed) line represents the 90\% (50\%) credible region.
}
\label{fig:GW190814}
\end{figure}

In \cref{fig:GW190412} and \cref{fig:GW190814}, we show the sky-localisation for the \zerospin model for GW190412 and GW190814 and compare these to the high- and low-latency results.
Here we see in detail that the \zerospin prior performs about as well as the low-latency localisation (finding an extra mode not present in the high-latency result, and overall a much broader area).
From this we conclude that optimizing the prior for black hole systems (which can exhibit significant spin and mass asymmetry) does not perform as well as a high-latency analysis with a complete prior specification. 

\section{Optimizing the choice of waveform model}
\label{sec:waveform}

In \cref{sec:result}, we demonstrated that for BNS and NSBH signals with small spins and moderate mass ratio's,  the sky localisation is unbiased by a \zerospinequalmass prior.
We analysed the data using the \IMRPhenomPvTwo \emph{waveform approximant} model which models a fully precessing binary black hole merger.
(Waveform approximants allows the generation of a predicted signal to within a few to tens of milliseconds. Their computation time is typically determined by their level of sophistication: approximants which better-model the underlying physics typically are slower to generate).
In this section, we aim to investigate if simpler waveform models can be used in place of more physically plausible models for localisation.

In \cref{tab:waveforms}, we provide a breakdown of the timing of the \IMRPhenomPvTwo waveform approximant, the \IMRPhenomPvTwoNRTidal (used to simulate signals in \cref{sec:simulated}), and the non-spinning \TaylorF \citep{TaylorF2} waveform which models only the inspiral of non-spinning point particles.
The \TaylorF is not as physically accurate as the IMRP waveforms which include the inspiral, merger, and ring-down (and tidal effects in the case of \IMRPhenomPvTwoNRTidal) of precessing binary systems.
This lack of physics translates into a likelihood which can be evaluated $\sim 30$\% faster than the \IMRPhenomPvTwo waveform.
If we are using a non-spinning and equal mass prior, then only remaining difference between \TaylorF and \IMRPhenomPvTwo is the physical modelling of the merger and ring-down.
However, the localisation is predominantly determined by triangulation from the inspiral signal.
This suggests a further optimization: use the \TaylorF waveform.

\begin{table}
    \centering
    \begin{tabular}{l|l|l}
         {Waveform approximant} & {per-likelihood} & {per-waveform}\\ 
         {} & {evaluation [ms]} & {evaluation [ms]} \\ 
         \hline
{\tt{IMRPhenomPv2\_NRTidal}} & $93 \pm 5$ & $53 \pm 4$ \\
{\tt{IMRPhenomPv2}} & $87 \pm 6$ & $47 \pm 4$ \\
{\tt{TaylorF2}} & $60 \pm 8$ & $13.3 \pm 0.7$ \\
    \end{tabular}
    \caption{Per-likelihood and per-waveform evaluation times.
    The per-likelihood captures the full cost of computing the likelihood in \bilby for simulated data lasting 128s.
    The difference between the per-likelihood and per-waveform values (which amounts to the \bilby data processing overhead) is $\sim 40$~ms.
    For the \TaylorF waveform, the data processing overhead is the dominant contribution to the per-likelihood evaluation.
    These timing apply only to the standard configuration (without any analytical marginalization).
    They can be drastically reduced by ROQ and hetrodyne methods as discussed in Section~\ref{sec:introduction}.
    All timings are bench-marked on an Intel Core i7-7820HQ CPU @ 2.90GHz.}
    \label{tab:waveforms}
\end{table}

\begin{figure}
    \centering
    \includegraphics[width=3.1in]{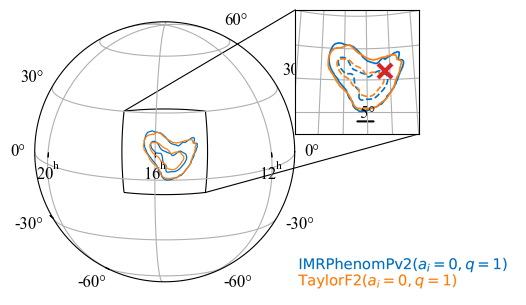}
    \caption{2D sky localisation of a simulated BNS system comparing the \IMRPhenomPvTwo and \TaylorF waveform models with identical \zerospinequalmass priors.
    The red ``cross'' indicates the simulated signal value.}
    \label{fig:waveform_skymap}
\end{figure}

To demonstrate that such a waveform optimization does not bias the result, we first look at a fiducial simulated signal.
We simulate a spinning BNS (simulation parameters: $q=0.7$, $a_1=0.04$, $a_2=0.01$, $\Lambda_1=1500$, $\Lambda_2=750$, and $d_{L}$=\SI{150}{Mpc}) in \SI{128}{s} of a two-detector network (HL) assuming O4 design-sensitivity noise.
The signal has an a simulated network optimal SNR of $\sim 18$.
We analyse the signal using \IMRPhenomPvTwo and \TaylorF waveforms under a \zerospinequalmass prior and plot the 2D sky localisation in Fig.~\ref{fig:waveform_skymap}.
This figure demonstrates that we do not see systematic difference between the two waveforms, despite their differing physical assumptions.
(The inferred luminosity distances show a similar level of agreement).

Both the \IMRPhenomPvTwo and \TaylorF analyses require a similar number of likelihood evaluation ($\sim 30\times10^{6}$), but the \TaylorF run had an overall wall time $30\%$ less than that of the \IMRPhenomPvTwo analysis. This confirms that the reduction in per-likelihood evaluations demonstrated in \cref{tab:waveforms} translate directly into wall time reductions.
For reference, we also analyse the data with the \IMRPhenomPvTwoNRTidal waveform and a full prior (i.e. using the full range of spins and tidal parameters).
Comparing the wall times, the \zerospinequalmass \TaylorF analysis is 60\% faster.

\begin{figure}
    \centering
    \includegraphics[width=.48\textwidth]{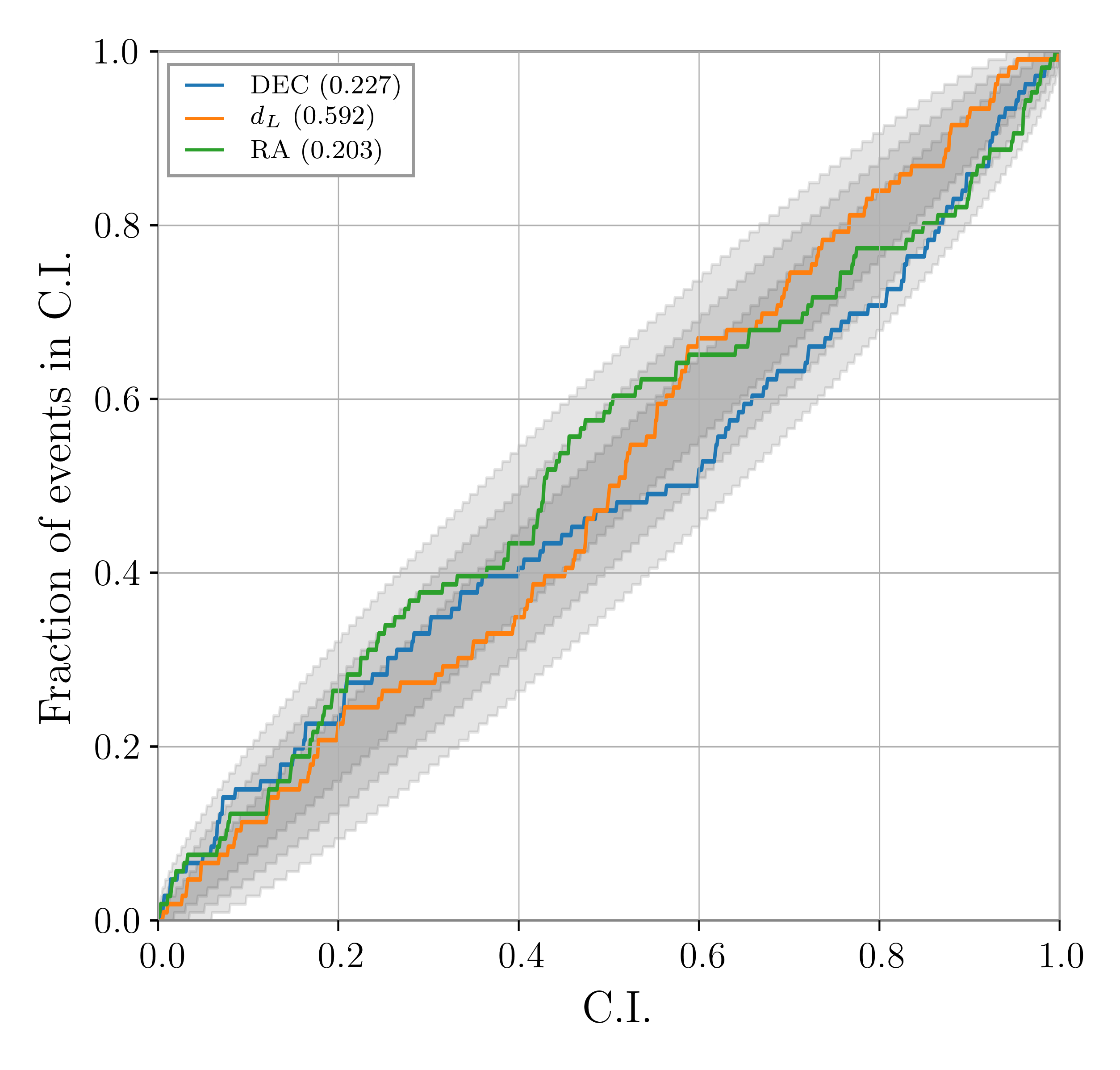}
    \caption{A parameter-parameter (PP) test of the 3D sky localisation parameters of 100 simulated BNS signal using the \zerospinequalmass prior. We simulate signals drawn from the astrophysical prior \cref{tab:priors} using the \IMRPhenomPvTwoNRTidal waveform model and analyse the data using the \TaylorF waveform model. The grey region indicates the 1, 2, and 3$\sigma$ confidence intervals. Individual p-values for each of the 3D localisation parameters are shown in legend.}
    \label{fig:pp_TF2}
\end{figure}

Finally, we verify that using the \TaylorF does not introduce a bias. We repeat the PP-test introduced in \cref{fig:pp_0S1Q}, but use the \TaylorF waveform approximant to analyse the data.
We report the results in \cref{fig:pp_TF2}.
Again, the 3D localisation remain inside the 3$\sigma$ uncertainty region. The individual p-value for DEC, $d_L$, and $\rm{RA}$, are 0.227, 0.592, and 0.203 respectively.

\section{Summary} \label{sec:summary}
We investigate optimization choices for the Bayesian prior and signal model used to produce sky localisation's from ground-based gravitational-wave observatories.
The rapid production of these sky localisation's is critical to aid in the search for possible counterparts.
For BNS systems (where we expect small spins and near equal mass), we demonstrate that a restrictive prior with zero spin and equal mass can reduce the number of required likelihood evaluations and hence wall time by $\sim 40\%$.
At the same time, we demonstrate that this optimized prior is unbiased, provided the signal does not have extreme spins or mass ratios.
We also demonstrate that using physically simpler waveform models provide equivalent sky-maps with up to a $\sim 30\%$ reduction in wall time.
Taken together, for BNS and moderate mass ratio NSBH systems optimized choices of the prior and waveform can reduce the wall time by up to 60\%.
This efficiency saving can be directly applied to the current stochastic sampling methods used in high-latency (e.g. by implementing the optimized prior in the \LALInference or \bilby samplers).
For BBH system, we demonstrate that the same prior optimizations do not apply: zero spin and equal mass assumptions produce poorer sky localisation.

We considered these optimization choices in the context of standard stochastic sampling.
As discussed in \cref{sec:introduction}, a number of new ideas are being developed which offer wall time speed-ups up to factors of a few hundred.
Prior and waveform optimization can be applied to both hetrodyning and brute-force parallelisation.
For ROQ-based methods, a speed-up is achieved by reducing the size of the basis (since it no longer needs to model unequal mass or spinning BNS), resulting in a speed-up of the basis itself.
For machine-learning based approaches, the optimizations described herein can be used to simplify the training set used at the learning stage.
Before the next observing run, we advocate for comparative head-to-head mock data challenges to ascertain which method is the fastest and most robust. This would include extending the studies of the optimized prior choices herein.

Whichever sampling method is used during the O4 observing run, we advocate that for BNS and NSBH systems, two localisation analyses be performed.
First, an optimized localisation which uses a zero-spin and equal-mass prior (and an inspiral-only waveform if applicable). Second, a complete localisation which uses a moderate low-spin and non-equal-mass configuration.
The optimized localisation can be produced in about half the time of the complete localisation. We anticipate that the two will differ only at the level of stochastic sampling.
Nevertheless, by running both analyses it can be assured that for system with significant spins and or highly asymmetric masses an updated localisation can be produced.

\section{acknowledgments}
We thank Soichiro Morisaki for useful feedback during the development of this work.
In analysing GW190425, we made use of the ROQ basis provided by \citet{baylor2019}.
This work was supported by the Australian Research
Council (ARC) Future Fellowship FT150100281 and
Centre of Excellence CE170100004 and also  
the National Natural Science Foundation of China under Grants Nos. 11633001
and 11920101003.
This work was performed on the OzSTAR national facility at Swinburne University of Technology. The OzSTAR program receives funding in part from the Astronomy National Collaborative Research Infrastructure Strategy (NCRIS) allocation provided by the Australian Government.
This material is based upon work supported by NSF’s LIGO Laboratory which is a major facility fully funded by the National Science Foundation.

\section{Data Availability}
This research has made use of data, software and/or web tools obtained from the Gravitational Wave Open Science Center (https://www.gw-openscience.org/ ), a service of LIGO Laboratory, the LIGO Scientific Collaboration and the Virgo Collaboration. LIGO Laboratory and Advanced LIGO are funded by the United States National Science Foundation (NSF) as well as the Science and Technology Facilities Council (STFC) of the United Kingdom, the Max-Planck-Society (MPS), and the State of Niedersachsen/Germany for support of the construction of Advanced LIGO and construction and operation of the GEO600 detector. Additional support for Advanced LIGO was provided by the Australian Research Council. Virgo is funded, through the European Gravitational Observatory (EGO), by the French Centre National de Recherche Scientifique (CNRS), the Italian Istituto Nazionale di Fisica Nucleare (INFN) and the Dutch Nikhef, with contributions by institutions from Belgium, Germany, Greece, Hungary, Ireland, Japan, Monaco, Poland, Portugal, Spain.

\bibliographystyle{mnras}
\bibliography{references}

\appendix
\section{Additional details on the PP test}
\label{sec:append}

\begin{figure}
\centering
\includegraphics[width=.48\textwidth]{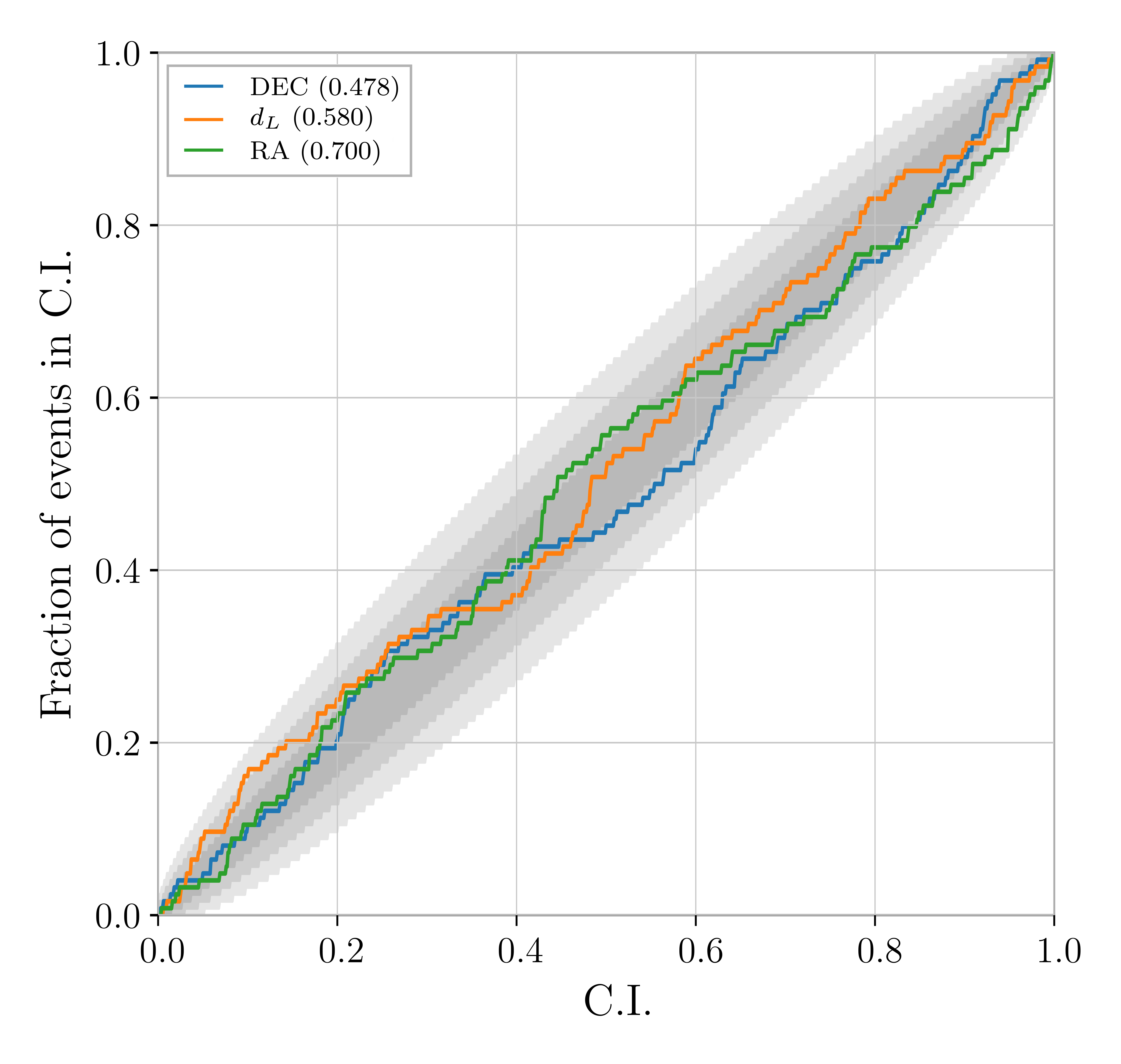}
\caption{A parameter-parameter (PP) test of the 3D sky localisation parameters of 100 simulated BNS signal using the \zerospinequalmass prior with a two-detector (H-L) network.}
\label{fig:pp_2det}
\end{figure}

\begin{figure}
\centering
\includegraphics[width=.48\textwidth]{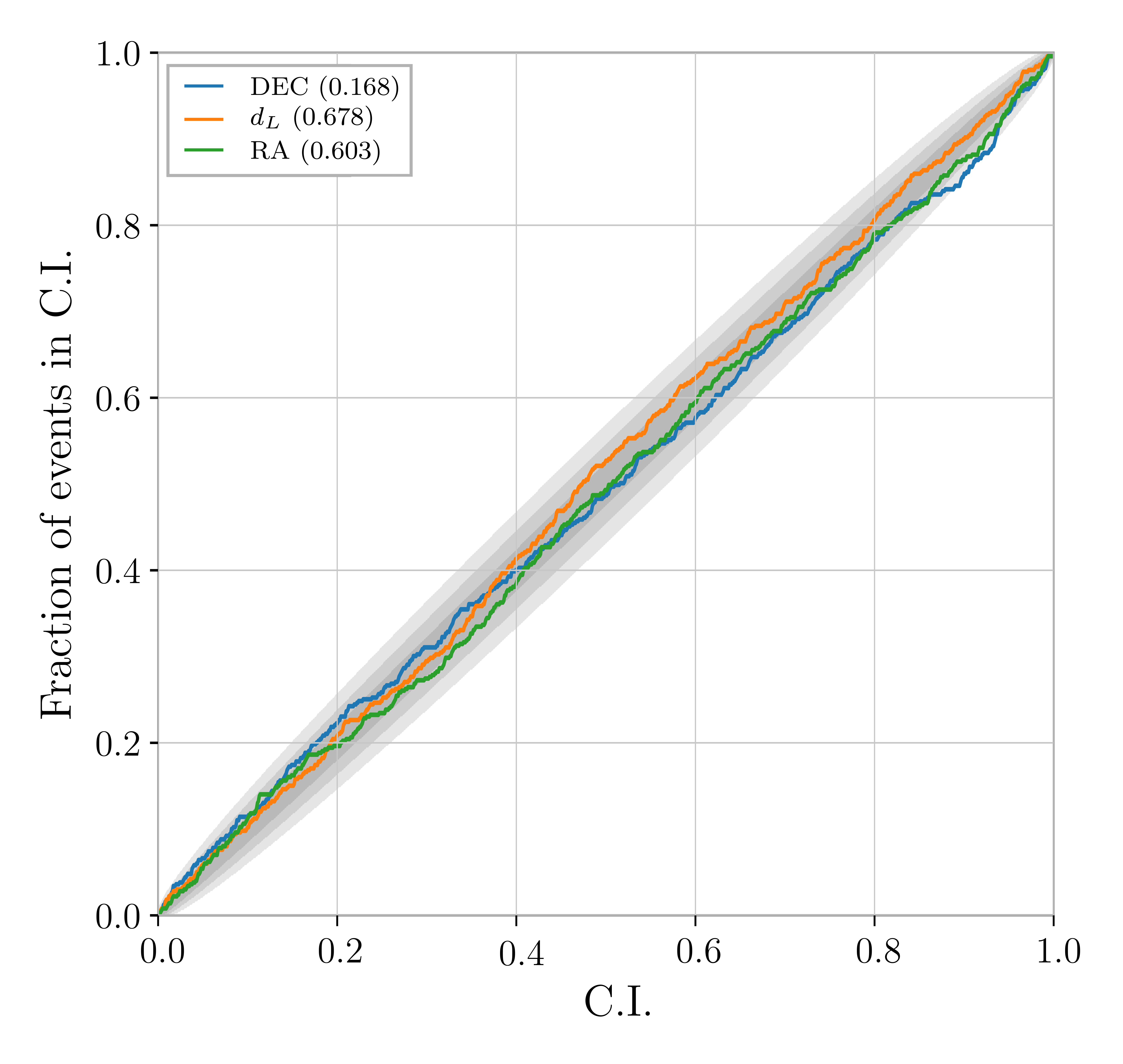}
\caption{As Fig. \ref{fig:pp_0S1Q} but for 500 injections.}
\label{fig:pp_500}
\end{figure}

{Here, we present additional details of the PP test we perform for our optimised choice of priors. In Fig. \ref{fig:pp_2det}, we show the PP plot for the \zerospinequalmass prior for a two-detector (H-L) network.
In Fig. \ref{fig:pp_500}, we show the PP plot for the \zerospinequalmass prior for the HLV detector network using 500 injections.
Furthermore, we list the p-values and the fractions of truth parameter values being out of the 90\% credible intervals for the \zerospinequalmass prior, for 50, 100, 300, and 500 injections.}

\begin{table}
  \centering
  \caption{The p-values and the fraction of the truth value out of 90 credible interval for DEC, $d_L$, RA with the 50, 100, 300, 500 simulated BNS injections.}
    \begin{tabular}{r|rrr|rrr}
    \hline\hline
          & \multicolumn{3}{c|}{p-values} & \multicolumn{3}{c}{out-of-90\%} \\
 \multicolumn{1}{c|}{N}  & \multicolumn{1}{c}{DEC} & \multicolumn{1}{c}{$d_L$} & \multicolumn{1}{c|}{RA} & \multicolumn{1}{c}{DEC} & \multicolumn{1}{c}{$d_L$} & \multicolumn{1}{c}{RA} \\
    \hline
    50    & 0.533 & 0.493 & 0.616 & 18\%  & 8\%   & 12\% \\
    100   & 0.122 & 0.551 & 0.710  & 13\%  & 12\%  & 9\% \\
    300   & 0.120  & 0.527 & 0.696 & 12\%  & 10\%  & 11\% \\
    500   & 0.168 & 0.678 & 0.603 & 13\%  & 9\%   & 11\% \\
    \hline\hline
    \end{tabular}
  \label{tab:comp_n}
\end{table}

\label{lastpage}
\end{document}